\input{aipcheck.tex}

\documentclass{aipproc}

\layoutstyle{6x9}

\SetInternalRegister\hbadness{8000} % pseudo latin isn't breaking very well :-)

\newcommand\doingARLO[2][]{%
  \ifx\mmref\undefined #1\else #2\fi
}

\begin{document}

\title{Status of the RHIC Spin Program}


\author{L.C. Bland}{
  address={Brookhaven National Laboratory, Upton, NY  USA},
  email={bland@bnl.gov}
}

% \copyrightholder{Acoustical Scociety of America}
\copyrightyear  {2004}

\begin{abstract}
The Relativistic Heavy Ion Collider (RHIC) at Brookhaven National
Laboratory has been developing the capability of accelerating, storing
and colliding high-energy polarized proton beams over the past several
years.  During this development phase, important first measurements of
cross sections and spin asymmetries for neutral pions produced in
polarized proton collisions at $\sqrt{s}$=200 GeV have been completed
by STAR and PHENIX, the two large collider experiments at RHIC.  This
contribution reports on progress of the RHIC spin program and
provides an outlook for the future.

\end{abstract}

\maketitle

\date{\today}

\section{Introduction}

To date, our knowledge of how the spin of the proton is
distributed amongst its quark and gluon constituents comes from
studies of deep-inelastic scattering (DIS) of polarized electrons and muons
from polarized proton, deuteron and $^3$He targets.  Global analyses
of this data ({\it e.g.}, Ref.~\cite{bb} and references to the data
therein) have provided a map of the Bjorken-$x$ dependence of the
quark (and anti-quark) polarization.  These studies have left us with
many puzzles, the most significant of which is how the proton gets its
spin from its constituents.  The spin physics program at the Relativistic
Heavy Ion Collider (RHIC) at Brookhaven National Laboratory
\cite{Bunce} is an ambitious effort to address the puzzles by
measuring spin asymmetries of particles produced in polarized proton
collisions at center of mass energy in the range 200 $\le \sqrt{s}
\le$ 500 GeV.  The object is to utilize polarized quarks as a probe of
the polarization of the constituents of the proton in scattering
processes that are described by perturbative QCD.  To achieve this
objective, it is necessary to establish that pQCD quantitatively describes
experimental cross sections and particle correlations at these
collision energies.

One of the long term objectives of the RHIC spin program is to determine
the Bjorken $x$ dependence of the helicity asymmetry distribution of
gluons within a longitudinally polarized proton, $\Delta G(x)$.
Completion of double helicity asymmetry ($A_{LL}$) measurements for
particles produced in different pseudorapidity ($\eta$) ranges and
as a function of transverse momentum ($p_T$) can result in a detailed map
of gluon polarization over a broad range of Bjorken $x$.  Inclusive meson
and jet production probes $\Delta G(x)$ through multiple partonic
subprocesses with contributions from a broad range of $x_{gluon}$ as specified by the
standard convolution integrals for hard scattering processes.
Di-hadron and di-jet production can provide additional constraints on
the kinematics, thereby changing the admixture of partonic
subprocesses and providing tighter constraints on the Bjorken $x$
values of the partons that participate in the hard scattering.  
Prompt photon production is expected to be dominated by
the QCD Compton process ($qg\rightarrow q\gamma$), thereby providing a
more selective probe of $\Delta G(x)$.  Further information can be
provided by the measurement of $A_{LL}$ for particle correlations,
such as $\gamma$+jet coincidences \cite{EPIC}.  At $\sqrt{s}$=500 GeV,
measurements of the parity violating single spin asymmetry $A_L$ for
$W^\pm$ production will result in the flavor separation of quark
polarization at large Bjorken $x$ and anti-quark polarization at
moderately small $x$, particularly when the daughter leptons from
$W^\pm$ decay are detected at large rapidity.  The unknown
transversity structure function can be probed via measurement of the
transverse spin correlation coefficient ($A_{TT}$) for inclusive jet
production. 

\section{How RHIC spin works}

High-energy electrons develop transverse polarization in a storage ring because
of a small spin flip probability in the sychrotron radiation they emit.
Muons are produced through the 
decays of pion beams, and so are naturally polarized because the weak
interaction violates parity.  To provide significant polarization for
intense proton beams at high energy, it is necessary to use atomic
methods at an ion source and then preserve the polarization through the
acceleration sequence.  The development of Siberian snakes
\cite{siberian} has enabled
the preservation of polarization while accelerating protons to high
energies through numerous depolarizing resonances in synchrotrons.  To
date, the highest energy polarized proton beam produced in a
synchrotron is at RHIC and is 100 GeV.  Previously, the highest energy
polarized proton collisions were studied using 200 GeV proton beams in
fixed target experiments.  The beams were produced by the
parity-violating decays of hyperons \cite{e704beam} and the total
center-of-mass energy for those experiments was $\sqrt{s}$=20 GeV.  At
RHIC, the polarized proton beams collide, enabling spin physics
experiments at collision energies 200$\le\sqrt{s}\le$500 GeV.  There
are plans to accelerate polarized proton beams with
energies greater than 200 GeV in the upcoming RHIC run as part of the
development of the capability for studying spin asymmetries for
$W^\pm$ production.  Exquisite care must be taken throughout the
entire acceleration chain to ensure that the polarization is
preserved.  Significant progress has been made in the development of
these capabilities in the first years of RHIC operations.

Also differing from polarized DIS experiments is the need to develop
methods to measure the polarization of high energy polarized protons.
Precision measurements of spin observables requires accurate
determination of the beam polarization.  For polarized lepton beams
used in DIS, quantum electrodynamics can be used to accurately predict
cross sections and spin observables for scattering processes.  Spin
dependent scattering processes involving polarized protons must be
calibrated by experiment since robust theoretical calculations of spin
observables at small momentum transfer, where interaction rates are
large, cannot be reliably made from first principles.

The RHIC polarimeters are based on measuring proton elastic scattering
from carbon at very small four momentum transfer, in the range 0.006
$\le -t \le$ 0.03 (GeV/c)$^2$ \cite{CNI}.  An analyzing power ($A_N$)
is expected because of the interference of the Coulomb spin-flip
amplitude and the spin-indepedent strong 
interaction amplitude; hence, the polarimetry process is referred to
as Coulomb-Nuclear Interference (CNI).  The observed $-t$ dependence
of the measured spin-dependent asymmetry, $P_{beam}\times A_N$,
suggests substantial contributions from the hadronic spin flip
amplitude, meaning that independent measurements are necessary to
determine the beam polarization ($P_{beam}$).  Precision measurements were
made for the first time at 100 GeV in RHIC run 4 (April through May, 2004) by
exploiting identical particle symmetries in polarized proton elastic
scattering from a polarized gas jet target \cite{bravar,hokada}.  The
contribution of the spin dependendent strong interaction amplitude to
small $|t|$ elastic scattering is of interest in its own right, and
can be extracted from spin observables measured at RHIC in either
fixed target experiments with the polarized gas jet target or with
carbon ribbon targets, or by studying elastic scattering of the
polarized colliding beams \cite{alekseev}, which is the object of the
pp2pp experiment at RHIC.

The first polarized proton collisions at $\sqrt{s}$=200 GeV were
achieved during RHIC run 2 (January through May, 2002).  Typical
collision luminosities were $\sim0.5 \times 10^{30}$ cm$^{-2}$s$^{-1}$ and the
average polarization was $\sim $16\%, inferred from assumptions
about the beam energy dependence of the CNI analyzing power \cite{trueman}.
Significant polarization loss in the RHIC injector resulted because the rate of
acceleration in the AGS was two times smaller than expected because of
equipment failure.  An integrated luminosity of 0.3 pb$^{-1}$ was
observed at the STAR interaction point.  Only vertical beam
polarization was available.

Polarized proton collisions at $\sqrt{s}$=200 GeV were also
achieved during RHIC run 3 (March through May, 2003).  Typical collision
luminosities observed with all subsystems operational at STAR were $3
\times 10^{30}$ cm$^{-2}$s$^{-1}$ and the beam polarization averaged
$\sim$25\%, based again on assumptions about the beam energy
dependence of the effective analyzing power of the polarimeter.  In
this run, helical dipole magnets located either side of PHENIX and
STAR were used to produce longitudinal polarization at the interaction
points.  Local polarimeters were developed by both experiments to
establish the absence of transverse polarization components for the
two beams at the interaction point concurrent with measurement of
spin-dependent asymmetries by the CNI polarimeters that are located in
parts of the ring where the beam polarization is vertical.  The
combined results from the local polarimeters and the CNI polarimeters
established that the colliding beams at STAR and PHENIX had
longitudinal polarization.  Integrated luminosities of 0.5 pb$^{-1}$
with vertical polarization and 0.4 pb$^{-1}$ with longitudinal
polarization were observed at the STAR interaction point.

Time was devoted during run 4 to accelerator development interleaved
with the polarization calibration experiment.  A new betatron tune for
RHIC was developed to reduce beam loss induced by the influence of one
beam on the other at the interaction points.  With unpolarized beam,
collision luminosities in excess of 10$^{31}$cm$^{-2}$s$^{-1}$ were
achieved during tests.  It was also established that beam polarization
could be preserved with the new tune of the accelerator.
Commissioning of a new warm-bore partial helical snake in the AGS was
successful.  At the end of the run there was a successful
demonstration of repeated stores in RHIC with an average collision
luminosity of $4 \times 10^{30}$ cm$^{-2}$s$^{-1}$ with average
polarization in both beams of $\sim$45\%.

For the immediate future, a long polarized proton run is planned for
the RHIC run starting in November, 2004.  Performance comparable to
that demonstrated during run 4 is expected to result in proton
collisions with an average polarization of 45\% and an average
integrated luminosity of 1 pb$^{-1}$ per week at a collision energy of
$\sqrt{s}$=200 GeV.  As discussed below, this should result in good
measurements of $A_{LL}$ for inclusive $\pi^0$ and jet production,
thereby providing sensitivity to possible gluon polarization.  A new
superconducting partial helical snake magnet for the AGS is planned to
be commissioned during this run.  It is anticipated that this device
will ultimately permit full transmission of the polarization from the ion source
into RHIC at the design intensity of $2\times10^{11}$ protons per
bunch.  Plans have been formulated to resolve the vacuum breakdown
issue within RHIC which presently limits luminosity.  When these
developments are complete, it is expected that design luminosity and polarization for $\vec{p}+\vec{p}$
collisions will be realized.

\section{What has been learned to date?}

\begin{figure}
\caption{Results for ({\it left}) the invariant cross section for neutral pions
produced at midrapidity by the PHENIX collaboration \cite{phenix_cross}
and ({\it right}) the invariant cross section for neutral pions
produced at large rapidity by the STAR collaboration \cite{star_cross}
in p+p collisions at $\sqrt{s}$=200 GeV in comparison with next-to-leading 
order perturbative QCD calculations.}
\includegraphics[height=0.5\textheight]{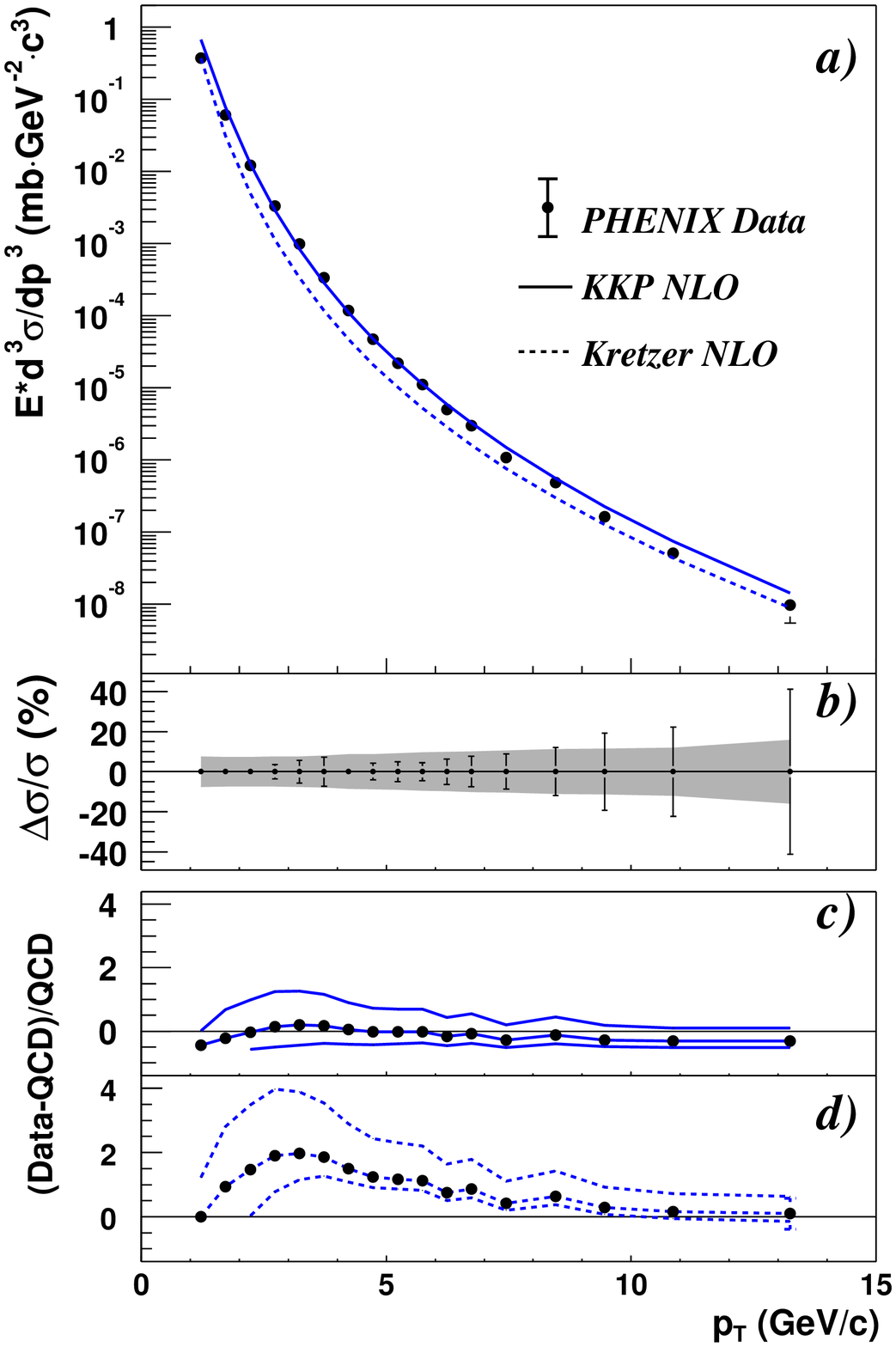}\includegraphics[height=0.42\textheight]{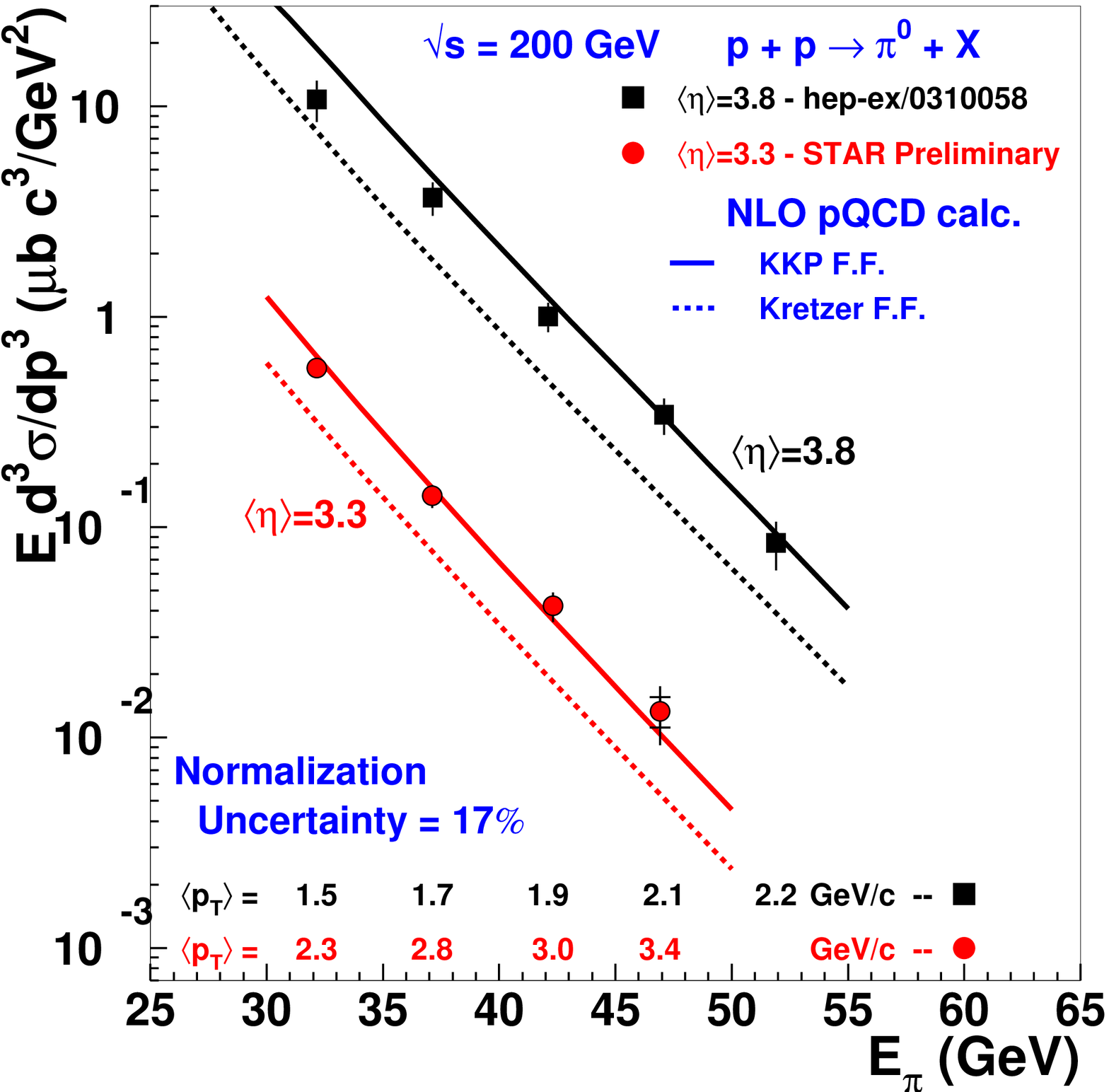}
\label{fig:a}
\end{figure}

The first polarized proton collisions at RHIC at $\sqrt{s}$=200 GeV
resulted in measurements of the cross section for neutral pion
production at both PHENIX and STAR (Fig.~\ref{fig:a}).  The results were for quite
different $\pi^0$ rapidities, with PHENIX reporting the $p_T$
dependence of the cross section near $\eta\approx$0 and STAR reporting
the Feynman $x$ dependence of the cross section at $<\eta>$=3.3 and
$<\eta>$=3.8.  These data have been compared with state of the art
perturbative QCD calculations \cite{nlo} computed at next-to-leading order.
The calculations use the CTEQ6M \cite{cteq} parton distribution
functions and the KKP \cite{KKP} and Kretzer \cite{kretzer}
fragmentation functions.

The theoretical calculations are found to quantitatively represent the
measured cross sections at $\sqrt{s}$=200 GeV over a broad range of
transverse momentum that extends down to $p_T \approx$ 1 GeV/c.
Agreement between calculation and data for transverse momenta
between $1\le p_T \le 10$ GeV/c had also been observed for inclusive
particle production at the Tevatron at much higher $\sqrt{s}$
\cite{KKP}.  The results from RHIC form part of a picture of the
pseudorapidity and collision energy dependence of $\pi^0$ production.
Comparison of NLO pQCD calculations to data obtained at lower
$\sqrt{s}$ \cite{bs04} shows a systematic tendency for more
quantitative agreement as $\sqrt{s}$ increases.  It is also the case
that the agreement extends over an increasingly broader range of $\eta$ as
$\sqrt{s}$ increases.  As shown in Fig.~\ref{fig:a}, at RHIC
collision energies, agreement between first principles calculations
and measured cross sections extends over a broad rapidity interval.
This agreement suggests that rapidity dependence of particle
production can be exploited to significantly alter the admixture of
$gg$, $qg$ and $qq^\prime$ subprocess contributions and to emphasize
partonic contributions from different regions of Bjorken $x$.

Understanding $\pi^0$ production with the electromagnetic calorimeters
in PHENIX and from STAR is also an important prerequisite for using
these devices for prompt photon detection.  We can anticipate results
for cross sections \cite{kokada} and spin observables for prompt photons in the near future.

In addition to inclusive cross sections, the first RHIC runs have
produced results for particle correlations.  At mid-rapidity,
measurements of azimuthal correlations of pairs of charged hadrons
produced at midrapidity have been reported for p+p, d+Au and Au+Au
collisions at $\sqrt{s_{NN}}$=200 GeV \cite{btob}.  For p+p collisions
the correlations have the pattern expected for a partonic scattering
origin to high-$p_T$ particle production; namely, a relative narrow
peak when the hadrons have a small azimuthal angle difference
(`near-side') ($\Delta\phi$) and a broader, but well developed, peak
when they are separated by $\Delta\phi=180^\circ$ (`away-side').
These studies have been extended to hadron pairs separated by large
rapidity intervals \cite{ogawa}.  Strong azimuthal correlations are
observed for $\pi^0-h^\pm$ hadron pairs separated by four units of
rapidity.  To date, comparisons of azimuthal correlations with theoretical
predictions have been limited.  One analysis \cite{vogelsangboer}
has established the need for multiple-soft gluon emissions to
understand the width and shape of the away-side 
di-hadron $\Delta\phi$ distribution.  It is fully anticipated that
di-hadron correlations will provide a wealth of information when
confronted against NLO pQCD calculations.  Furthermore, future
measurements of transverse and longitudinal spin observables for
di-hadron, di-jet and $\gamma$+jet events will lead to many
interesting results.

\begin{figure}
\caption{The first spin asymmetries from RHIC from $\vec{p}+\vec{p}$ collisions at $\sqrt{s}$=200 GeV.
{\it (Left)} The Feynman $x$ dependence of the analyzing power 
for large rapidity $\pi^0$ production in comparison to QCD model expectations.  
{\it (Right)}  The $p_T$ dependence of the longitudinal spin asymmetry, $A_{LL}$ for
$\pi^0$ production at midrapidity.}
\includegraphics[height=0.34\textheight]{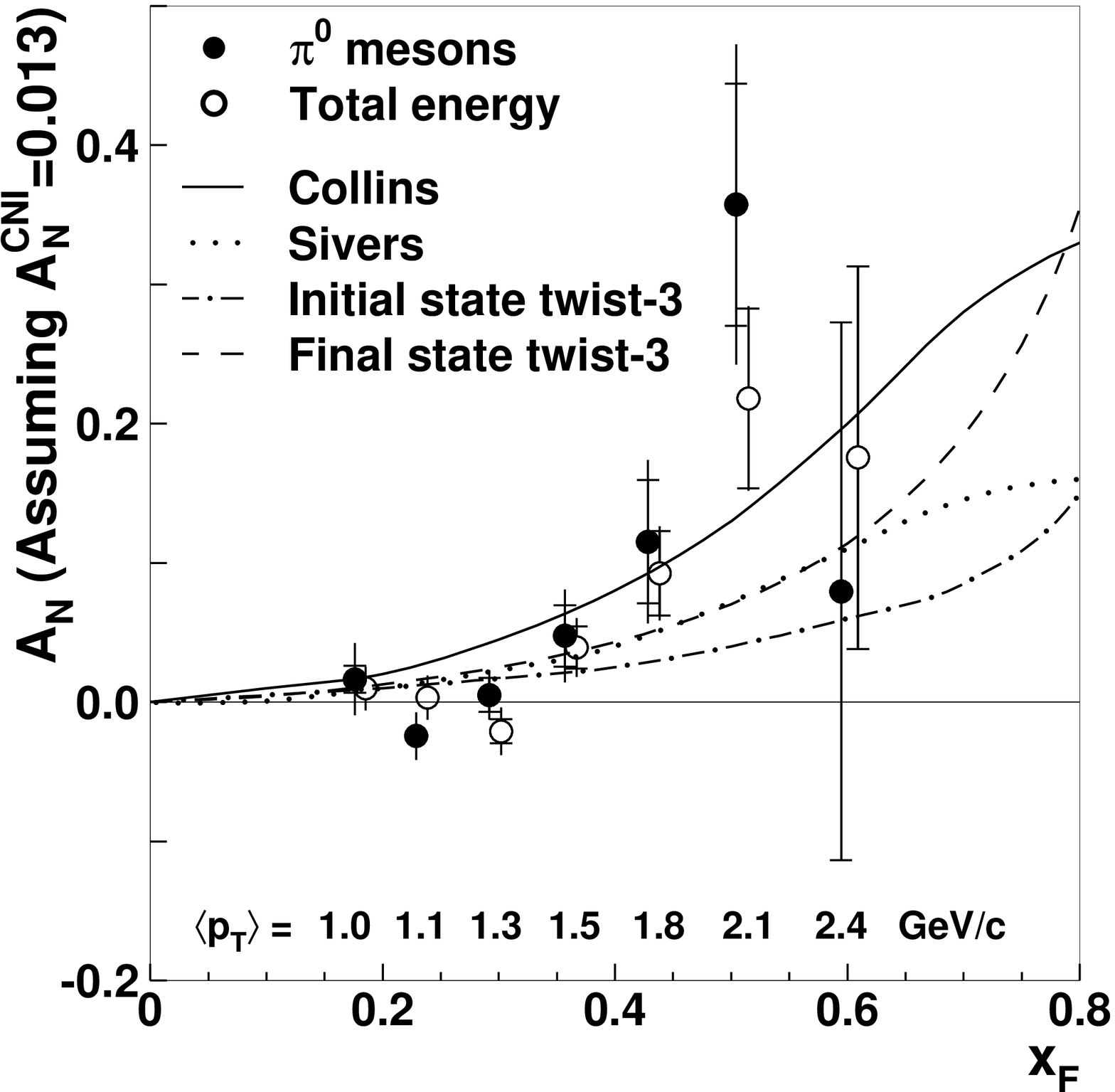}\includegraphics[height=0.4\textheight]{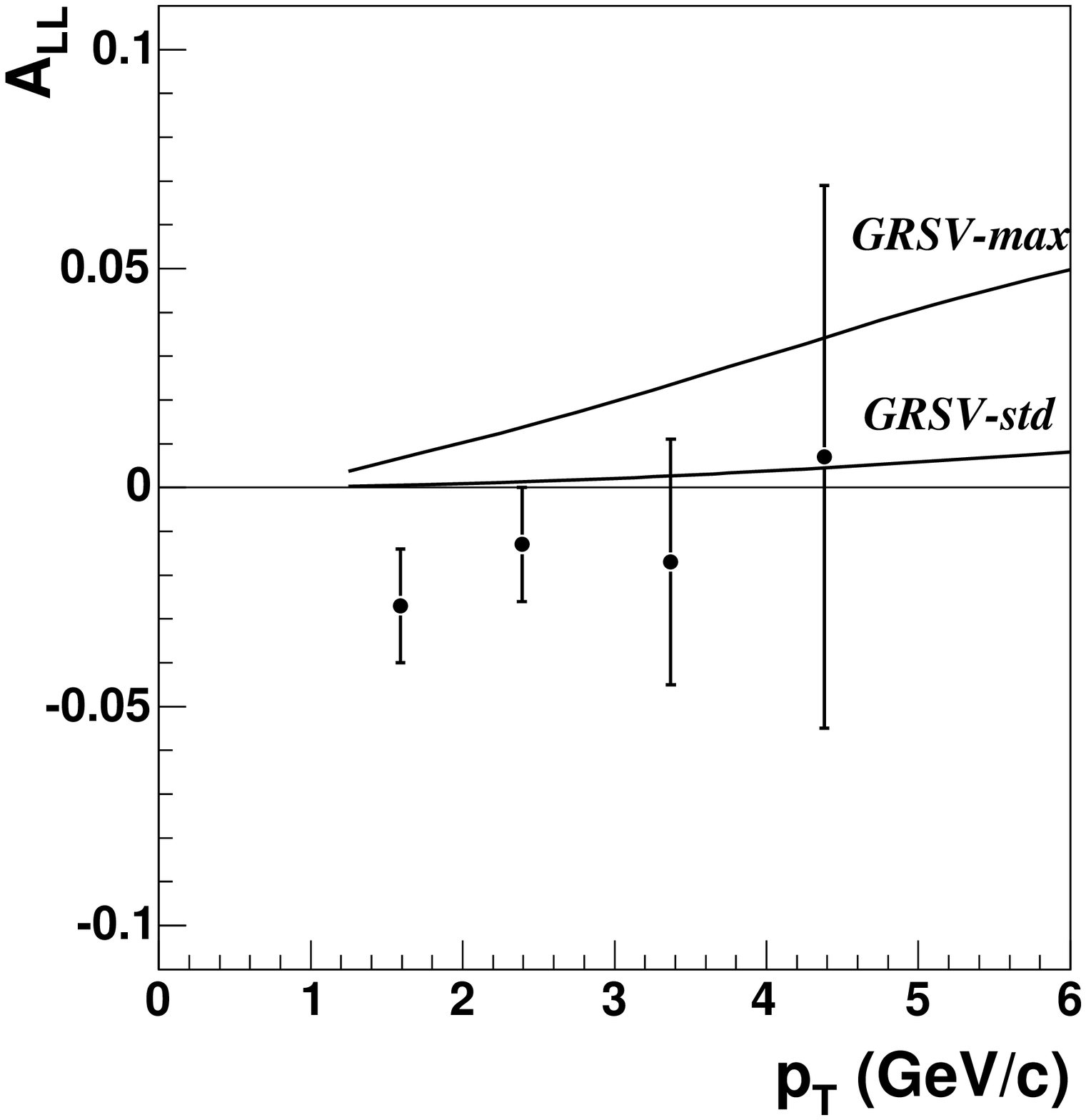}
\label{fig:b}
\end{figure}

The first measurements of spin observables in $\pi^0$ production have
been made as RHIC develops the polarization and intensity of the
colliding proton beams.  For $\pi^0$ produced at large rapidity, a
large analyzing power has been observed in transversely polarized
proton collisions at $\sqrt{s}$=200 GeV \cite{star_cross} (left side
of Fig.~\ref{fig:b}).  The pQCD picture that describes the unpolarized
cross section in these kinematics identifies the dominant partonic
process as large Bjorken-$x$ quarks (from the polarized proton)
interacting with soft gluons from the other proton.  At large $x$,
global analyses have established the quarks are highly polarized
within a longitudinally polarized proton.  If transversity behaves in
a similar fashion, then the large $A_N$ could result from highly
polarized quarks that undergo a spin- and transverse-momentum ($k_T$)
dependent fragmentation into $\pi^0$ (Collins effect)
\cite{collins,anselminocollins}.  It could also be that the quarks
involved in the particle production are from a spin- and
$k_T$-dependent distribution function associated with the polarized
proton (Sivers effect) \cite{sivers,anselminosivers}.  There are also
twist-3 mechanisms that can give rise to the observed $A_N$
\cite{highertwist}.  Further experiments are required to disentangle
these mechanisms.  Possibilities include the measurement of the spin
dependence of azimuthally correlated hadron pairs ({\it e.g.},
$\pi^0$-$h^\pm$).  Extension of these measurements to negative $x_F$
could isolate contributions from the Sivers function associated with
gluons, since gluons cannot contribute to transversity within the
proton, thereby eliminating the Collins effect.  It would also be
useful to separately study the $x_F$ and $p_T$ dependence of $A_N$.
We can also expect results from the BRAHMS experiment for the $x_F$
dependence of $A_N$ for charged pions \cite{videbaek}.

The $A_N$ results for $\vec{p}+\vec{p}$ collisions at $\sqrt{s}$=200
GeV have an $x_F$ dependence quite similar to that observed at
$\sqrt{s}$=20 GeV \cite{e704} and at lower energies
\cite{lowerenergy}.  One of the challenges is to obtain a robust
theoretical understanding of the dynamics responsible for this
phenomena at energies where NLO pQCD provides a good framework for
understanding the cross section and at lower energies where the theory
fails to explain data \cite{bs04}.

The right side of Fig.~\ref{fig:b} shows the first measurment of
$A_{LL}$ for $\pi^0$ production at RHIC, recently reported by the
PHENIX collaboration \cite{phenix_all}.  Data from midrapidity $\pi^0$
and jet production promise to provide insight into gluon
polarization.  First measurements of $A_{LL}$ for midrapidity $\pi^0$
production in $\vec{p}+\vec{p}$ interactions at $\sqrt{s}$=20 GeV were
made by the E704 collaboration \cite{e704all}.  The first $A_{LL}$
measurement at RHIC at $\sqrt{s}$=200 GeV is based on an integrated
luminosity of only 0.22 pb$^{-1}$ with average beam polarization of
27\%.  Even with these limited statistics the $\pi^0$ $A_{LL}$
measurement has accuracy comparable to the uncertainty on gluon
polarization deduced from inclusive DIS measurements, as reflected by
the two curves in the figure.  The GRSV-max curve is computed
\cite{jager} assuming 100\% gluon polarization at the input scale,
$Q_0^2$=0.6 GeV$^2$.  The GRSV-std curve represents the best global
fit \cite{gluck} to the inclusive polarized DIS data.  Given the
improved performance of polarized proton operations at RHIC, there is
promise for at least an order of magnitude improvement in the
statistical accuracy of $A_{LL}$ in the upcoming RHIC run.

What does this mean for gluon polarization?  Inclusive observables are
computed as a convolution of distribution functions, hard scattering
cross section and fragmentation functions.  Consequently, $\Delta
G(x)$ must be extracted from a global theoretical analysis, in a
manner similar to that used for unpolarized parton distribution
functions.  The midrapidity data probes Bjorken $x$ values that are on
average equal to $x_T = 2 p_T / \sqrt{s}$.  Tests of the resulting
$\Delta G(x)$ can be made by measurements of the rapidity dependence of
$A_{LL}$ for $\pi^0$ production; midrapidity inclusive jet production,
that would eliminate the fragmentation function from the convolution
integral; and di-hadron correlations, that can narrow the range of
contributions from the distribution functions depending on the
kinematics of the hadron pair.  Measurements of $A_{LL}$ for prompt
photon production are expected in the future, when the luminosity and
polarization of the RHIC beams are further developed.

\section{Summary and Outlook}

RHIC is the world's first polarized proton collider.  The first
polarized proton collisions have resulted in cross section data that
show fixed-order perturbative QCD is a robust framework for
understanding inclusive particle production at $\sqrt{s}$=200 GeV.
The first spin asymmetry measurements have begun for inclusive $\pi^0$
production.  The analyzing power at large $x_F$ is found to be large.
Additional measurements are needed to disentangle contributions from
different mechanisms.  The first measurements of the double helicity
asymmetry ($A_{LL}$) for midrapidity $\pi^0$ production have been
reported.  The expectations are that significantly more precise
measurements can be completed in the upcoming RHIC run.  The new data
will address whether gluon polarization makes significant
contributions to the spin of the proton.  Future measurements of
$A_{LL}$ for direct photon production will become feasible as the
luminosity and polarization at RHIC improves.  Future runs at
$\sqrt{s}$=500 GeV will measure parity violating spin asymmetries for
$W^\pm$ production.

\begin{theacknowledgments}
The progress in the RHIC spin program has been made by a very large
group of people.  This report is a summary of their work.  Groups involved
are the Collider-Accelerator Department at BNL, with primary
responsibility for developing the capability to accelerate, store and
collide polarized proton beams and the experiment collaborations; STAR
and PHENIX, with primary responsibility for building, operating and
analyzing the results from the collider detectors; and the polarimetry
group that have developed techniques for measuring the beam
polarization.  The RHIC spin project has received important support
from the the Department of Energy, the RIKEN institute and the National
Science Foundation.  

I would like to thank G. Bunce, A. Ogawa and G. Rakness 
for their careful reading of this manuscript.
\end{theacknowledgments}

% choose bibtex style depending on layout style and options used in
% sample:

\bibliographystyle{aipproc}

\begin{thebibliography}{1}

\bibitem{bb}
  J.~Bl\"umlein and H.~B\"ottcher, Nucl. Phys. {\bf B636},225 (2002).

\bibitem{Bunce}
  Gerry Bunce, Naohito Saito, Jacques Soffer and Werner Vogelsang, Ann.\
  Rev.\ Nucl.\ Part.\ Sci.\ {\bf 50}, 525 (2000).

\bibitem{EPIC}
  L.~C.~Bland, in {\it Physics with a High Luminosity Polarized Electron
  Ion Collider}, eds. L.~C.~Bland, T.~Londergan and A.~Szczepaniak
  (World Scientific, Singapore, 2000).  Also available at
  hep-ex/9907058.


\bibitem{siberian}
  Ya.S~Derbenev and AM. Kondratenko, Sov.\ Phys.\ JETP\
  {\bf 37}, 968 (1973); Ya.S~Derbenev {\it et al.}, Part.\ Acc.\ {\bf
  8}, 115 (1978). 

\bibitem{e704beam}
  D.~P.~Grosnick {\it et al.}, Nucl. Instr. Meth. {\bf A290} (1990) 269.

\bibitem{CNI}
  J.~Tojo {\it et al.}, Phys. Rev. Lett. {\bf 89} (2002) 052302;
  O.~Jinnouchi {\it et al.}, AIP Conf. Proc. {bf 675} (2003) 817.

\bibitem{bravar}
  A.~Bravar {\it et al}, 16th Int. Spin Physics Symposium (SPIN 2004),
  to be published in the proceedings.

\bibitem{hokada}
  H.~Okada {\it et al}, 16th Int. Spin Physics Symposium (SPIN 2004),
  to be published in the proceedings.

\bibitem{alekseev}
  I.~Alekseev {\it et all}, 16th Int. Spin Physics Symposium (SPIN
  2004), to be published in the proceedings.
\bibitem{trueman}
  T.~L.~Trueman, hep-ph/0203013.

\bibitem{nlo}
  F.~Aversa {\it et al.}, Nucl.\ Phys.\ 
  {\bf B327} (1989) 105;
  B.~Jager {\it et al.}, Phys.\ Rev.\ D\ {\bf 67} (2003) 054005; 
  D.~de~Florian, {\it ibid.}\ {\bf 67} (2003) 054004.

\bibitem{cteq} 
  J.~Pumplin {\it et al.}, J.\ High Energy Phys.\ 
  {\bf 07} (2002) 012.

\bibitem{KKP}
  B.\,A.~Kniehl {\it et al.}, Nucl.\ Phys.\ {\bf B597} (2001)
  337.

\bibitem{kretzer}
  S.~Kretzer, Phys.\ Rev.\ D\ {\bf 62} (2000) 054001.

\bibitem{phenix_cross}
  S.~S.~Adler, et al. (PHENIX collaboration), Phys. Rev. Lett. {\bf 91} (2003) 241803.

\bibitem{star_cross}
  J.~Adams, et al. (STAR collaboration), Phys. Rev. Lett. {\bf 92}
  (2004) 171801.

\bibitem{bs04}
  C.~Bourrely and J.~Soffer, 
  Eur. Phys. J. C {\bf 36} (2004) 371 and {\mbox hep-ph/0311110}.

\bibitem{kokada}
  K.~Okada {\it et al.} (PHENIX collaboration), 16th Int. Spin Physics
  Symposium (SPIN 2004), to be published in the proceedings.

\bibitem{btob}
  C.~Adler {\it et al.} (STAR collaboration), Phys. Rev. Lett. {\bf
  91} (2003) 172302; J.~Adams {\it et al.} (STAR collaboration),
  Phys. Rev. Lett. 91 (2003) 072304.

\bibitem{ogawa}
  A.~Ogawa, contribution to the 12th International Workshop on Deep
  Inelastic Scattering (to be published), {\mbox nucl-ex/0408004}.

\bibitem{vogelsangboer}
  Daniel Boer and Werner Vogelsang, Phys.Rev. D69 (2004) 094025.

\bibitem{collins}
  J.~Collins, Nucl.~Phys.~{\bf B396} (1993) 161.

\bibitem{anselminocollins} 
  M.~Anselmino, M.~Boglione, and F.~Murgia,
  Phys.\ Rev.\ D\ {\bf 60}, 054027 (1999).

\bibitem{sivers}
  D.~Sivers, Phys.~Rev.~D {\bf 41} (1990) 83; {\bf 43} (1991) 261.

\bibitem{anselminosivers} 
  M.~Anselmino, M.~Boglione, and F.~Murgia,
  Phys.\ Lett.\ B\ {\bf 362}, 164 (1995); 
  M.~Anselmino and F.~Murgia, {\it ibid.}\ {\bf 442}, 470 (1998).

\bibitem{highertwist}
  A.~Efremov and O.~Teryaev, Phys.~Lett.~{\bf 150B} (1985) 383;
  J.~Qiu and G.~Sterman, Phys.~Rev.~D {\bf 59} (1998) 014004;
  Y.~Koike, AIP Conf. Proc. {\bf 675} (2003) 449.

\bibitem{videbaek}
  F.~Videbaek {\it et al.} (BRAHMS collaboration), 2004 Fall Meeting
  of the Division of Nuclear Physics.

\bibitem{e704}
  B.\,E.~Bonner {\it et al.}, Phys.\ Rev.\ Lett.\ {\bf
  61} (1988) 1918; D.\,L.~Adams {\it et al.}, Phys.\ Lett.\ B {\bf 261} (1991)
  201; {\bf 264} (1991) 462; Z.\ Phys.\ C {\bf 56} (1992) 181; 
  A.~Bravar {\it et al}, Phys. Rev. Lett. {\bf 77} (1996) 2626.

\bibitem{lowerenergy}
  R.\,D.~Klem {\it et al.}, Phys.\ Rev.\ Lett.\
  {\bf 36} (1976) 929; W.\,H.~Dragoset {\it et al.}, Phys.\ Rev.\
  D\ {\bf 18} (1978) 3939; S.~Saroff {\it et al.}, Phys.\ Rev.\ Lett.\
  {\bf 64} (1990) 995; B.\,E.~Bonner {\it et al.}, Phys.\ Rev.\ D\
  {\bf 41} (1990) 13; K.~Krueger {\it et al.}, Phys.\ Lett.\ B\ {\bf
  459} (1999) 412; C.\,E.~Allgower {\it et al.}, Phys.\ Rev.\ D\
  {\bf 65} (2002) 092008.


\bibitem{phenix_all}
  S.~S.~Adler, et al. (PHENIX collaboration), Phys. Rev. Lett. {\bf 93} (2004) 202002.

\bibitem{e704all}
  D.~L.~Adams {\it et al.}, Phys. Lett. {\bf B261} (1991) 197.

\bibitem{jager}
  B.~J\"{a}ger {\it et al.}, Phys. Rev. D {\bf 67} (2003) 054005.

\bibitem{gluck}
  M. Gl\"{u}ck {\it et al.}, Phys. Rev. D {\bf 63} (2001) 094005.

\end{thebibliography}

\end{document}